\documentclass[proof]{WileyASNA-v2}

\articletype{Original Article}

\received{X X 2020}
\revised{X X 2020}
\accepted{X X 2020}

\begin{document}

\title{Time series of optical spectra of Nova V659 Sct}

\author[1]{Dennis Jack*}
\author[1,2]{Klaus-Peter Schr\"oder}
\author[1]{Philippe Eenens}
\author[3]{Uwe Wolter}
\author[3]{Jos\'e Nicol\'as Gonz\'alez-P\'erez}
\author[3]{J\"urgen H. M. M. Schmitt}
\author[3]{Peter H. Hauschildt}
\authormark{D. Jack et al.}

\address[1]{\orgdiv{Departamento de Astronom\'\i{}a}, \orgname{Universidad de Guanajuato}, \orgaddress{\state{Guanajuato}, \country{Mexico}}}
\address[2]{\orgdiv{Sterrewacht Leiden}, \orgname{Universiteit Leiden}, \orgaddress{\state{Leiden}, \country{Netherlands}}}
\address[3]{\orgdiv{Hamburger Sternwarte}, \orgname{Universit\"at Hamburg}, \orgaddress{\state{Hamburg}, \country{Germany}}}

\corres{*Dennis Jack, \email{dennis.jack@ugto.mx}}

\presentaddress{Departamento de Astronom\'\i{}a, Universidad de Guanajuato, A.P.~144, 36000 Guanajuato, GTO, Mexico}

\abstract{With our robotic 1.2~m TIGRE telescope, we were able to obtain eight optical spectra with intermediate resolution 
($R\approx 20,000$) of the Nova V659 Sct during different phases of its outburst.
We present a list of the lines found in the Nova spectra. The most common features are H~I, O~I, Na~I, Fe~II and Ca~II. 
Studying the spectral evolution of the strong features we found that the absorption 
features move to higher expansion velocities before disappearing and the emission features show (different) asymmetries.
Thanks to the intermediate spectral resolution we identified and analysed the interstellar medium absorption features present in the spectra. 
We detected atomic absorption features of Na~I and Ca~II. The sodium D lines
show more complex substructures with three main absorption features at around a velocity of $-10$, $30$ and $85$ km~s$^{-1}$. 
We identified several DIBs in the Nova V659 Sct spectra and determined their velocities and equivalent widths.}


\fundingInfo{DFG, CONACYT}

\maketitle

\footnotetext{\textbf{Abbreviations:} }

%
% Introduction
%
\section{Introduction}

The occassional appearance of "new stars" (Stella Nova) in the sky are 
bright outbursts known today as classical novae which are the result of the accretion process in a close binary
system consisting of a white dwarf (WD) and its main sequence or evolved companion star. 
The WD accumulates hydrogen from the companion star on its surface which eventually ignites and
explodes in a thermonuclear runaway. During this outburst 
the brightness of the WD increases by several magnitudes and one can observe these events as classical novae
with their typical light curves and spectral evolution
(see \citet{payne64,gallagher78,book08,bode10} or \citet{starrfield16} for reviews).

The galactic Nova V659 Sct (Nova Scuti 2019, AT 2019tpb, ASASSN-19aad, TCP J18395972-1025415) was discovered on October 29.058 UT \citep{disc}
by the ASAS-SN survey \citep{ASAS1,ASAS2} in the constellation Scutum. 
An independent discovery was reported by \citet{cbet4690} a few hours later.
Nova V659 Sct was quickly identified as a classical nova in its early stages of the outburst \citep{class}.
This nova had a $V$~band brightness of 8.4~mag during maximum optical light, a bright event which could be observed also
by small and medium-sized telescopes. However, its position close to the Sun in the sky made it
somewhat difficult to observe. 
Optical spectra were reported by \citet{opt} showing the typical lines for a classical nova with the common P-Cygni profiles.
Observations with the SWIFT telescope were performed \citep{swift}. The nova outburst was detected by instruments, both, in X-rays and the UV wavelength range. 
\citet{swift} determined that the nova was affected by galactic reddening of about $E(B-V)=0.9$.

Our robotic 1.2~m TIGRE telescope is an ideal instrument for the observation of time series of bright nova and even supernova events, like SN~2014J \citep{jack15b}.
Nova V659 Sct is the third nova that we could observe with our intermediate resolution optical spectrograph.
So far, we have obtained time series of the two classical Novae V339 Del \citep{novadel} and V5668 Sgr \citep{novasgr}.

The paper is structured as follows.
We will present in Section~2 details about the observations of the optical spectra of Nova V659 Sct that we obtained
with the TIGRE telescope.
An analysis of the observed spectra including a line identification, also for interstellar absorption features, is
presented in Section~3. We will close the presentation of our work with a summary in Section~4.

%
%  Observations 
%
\section{Observations of Nova V659 Sct}

\subsection{$V$ band light curve}
\begin{figure}
\begin{center}
 \resizebox{\hsize}{!}{\includegraphics{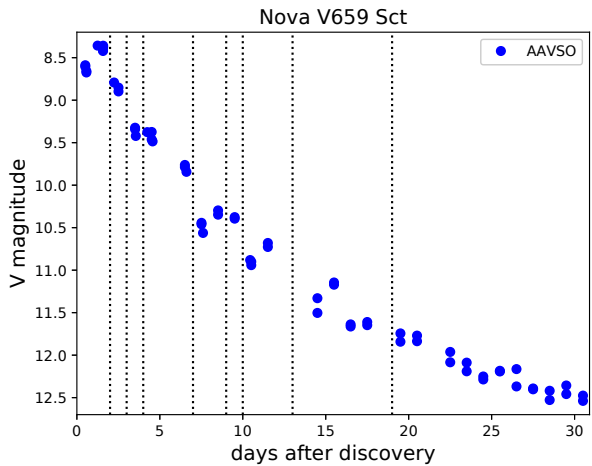}}
\caption{The $V$~band light curve of Nova V659~Sct. The data points were taken from AAVSO observations. Vertical lines mark TIGRE observations.}
\label{fig:lc}
\end{center}
\end{figure}
In Figure~\ref{fig:lc}, we present the $V$~band light curve of Nova V659~Sct using the prevalidated observations published by the AAVSO on its website\footnote{\url{https://www.aavso.org/}}. 
The position of the Nova in the constellation of Scutum was close to the Sun on the sky, and after about 30 days the Nova was not observable any more.
Nova V659 Sct was discovered before its the maximum optical light, which was reached around October 31, 2019. After the maximum the brightness dropped about 4 magnitudes over the
following 30 days.
The vertical lines in Figure~\ref{fig:lc} mark the observation times where we obtained optical spectra with the TIGRE telescope. 
We were able to take the first spectrum about two days after the discovery during the maximum phase of the $V$~band light curve.
Our last spectrum was obtained about 19 days after discovery because after 
that the Nova was too close to the Sun on the sky which did not allow us to have sufficient exposure time.

We applied the maximum magnitude versus rate of decline (MMRD) relationship of \citet{dellavalle} to estimate the distance to Nova V659~Sct.
Using $E(B-V)=0.9$ \citep{swift} and assuming $R_V=3.1$ we obtained an extinction
of $A_V=2.8$~mag. From the $V$ band light curve we determined its maximum brightness to have a value of $m_V=8.35$~mag. 
It is not possible to determine the time for the decline of 2 magnitudes very precisely because of few measurements and large scatter in the light curve.
We assumed a value of $t_2=7.5$~days. 
Applying the MMRD relation we obtained an absolute visual magnitude for Nova V659~Sct of $M_V=-8.8$~mag and, thus, a distance of 7.5~kpc to it. 
In order to verify this result we looked at the 3D Dust Mapping\footnote{\url{http://argonaut.skymaps.info/}} \citep{green19}. 
In the direction of Nova V659~Sct, the distance of 7.5~kpc correspondes to an extinction of $E(g-r)=0.96$, 
which roughly coincides with the before mentioned $E(B-V)=0.9$.

\subsection{TIGRE observations}

The observations were carried out with the TIGRE (Telescopio Internacional 
de Guanajuato Robotico Espectroscopico) telescope at
the observatory of La Luz located close to the town of Guanajuato in Central Mexico.
TIGRE is a robotic telescope with a mirror of an aperture of 1.2~m.
The HEROS (Heidelberg Extended Optical Range Spectrograph) echelle spectrograph has an
intermediate resolution of about $R\approx 20,000$
and obtains optical spectra in the wavelength range from about 3,800 to 8,800 \AA, 
having a small gap of about 120~\AA\ between the two arms around 5,800~\AA. 
Thus, the observed spectra are divided into two channels (blue and red). All observations and the data reduction are performed
automatically, which includes the barycentric correction.
For more technical details about the TIGRE telescope, consult \citet{schmitt14}.

Due to weather conditions we could obtain the first spectrum during the night of October 31, 2019. 
In total, we obtained eight optical spectra of Nova V659 Sct
during different phases of the spectral evolution. 
Fortunately, we were able to take three spectra during consecutive nights.
Our last spectrum was taken on November 17, about 19 days after the discovery.
The details of our observation campaign can be found in Table~\ref{obs_log}.

\begin{figure*}
\begin{center}
\includegraphics[width=\textwidth]{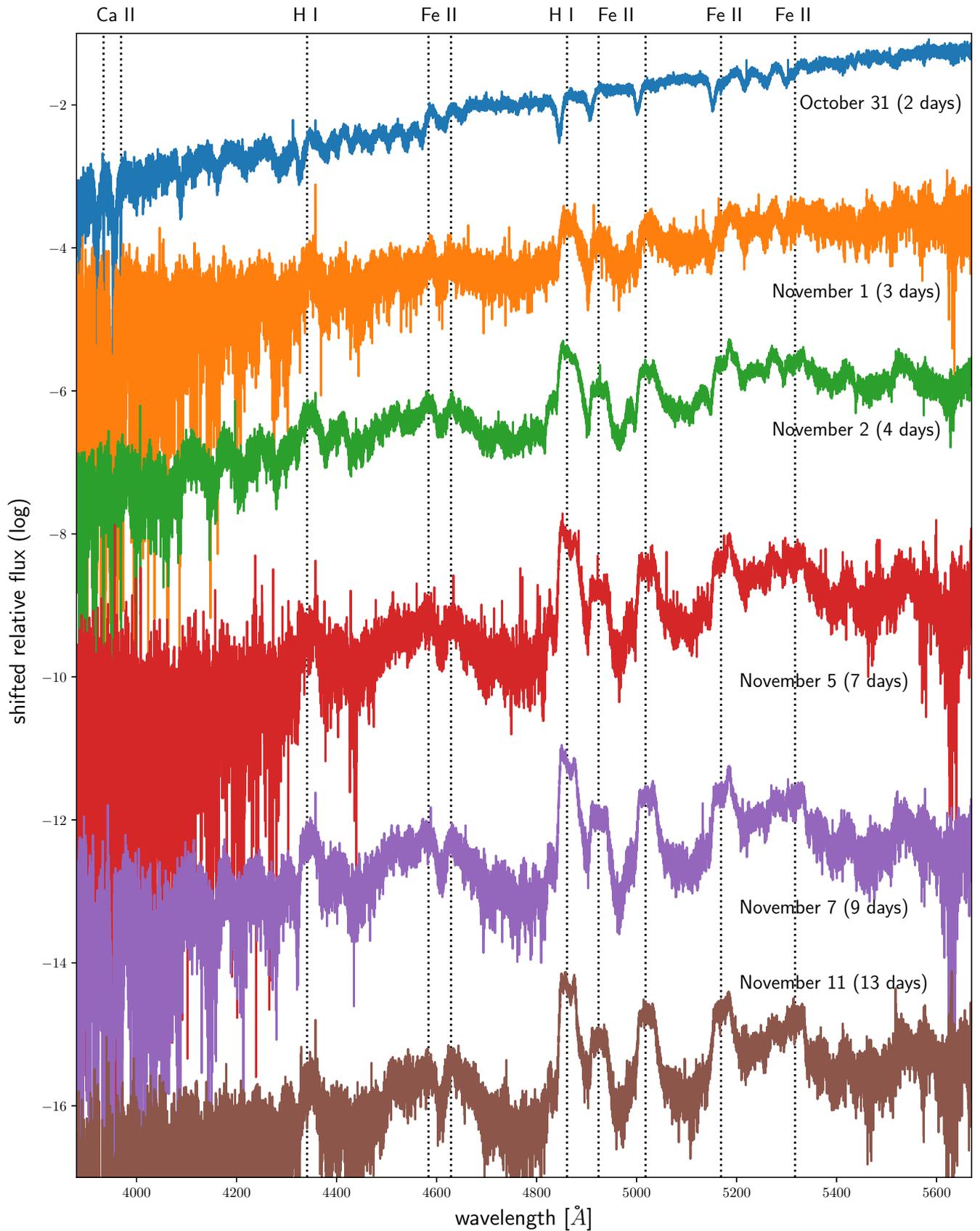}
\caption{Six spectra of Nova V659 Sct observed in the blue channel
of the HEROS spectrograph. Dates are of 2019 and the corresponding days
after discovery are given in brackets. The part below $4400$~\AA\ suffers from very low S/N.}
\label{fig:all_specs_B}
\end{center}
\end{figure*}

\begin{figure*}
\begin{center}
\includegraphics[width=\textwidth]{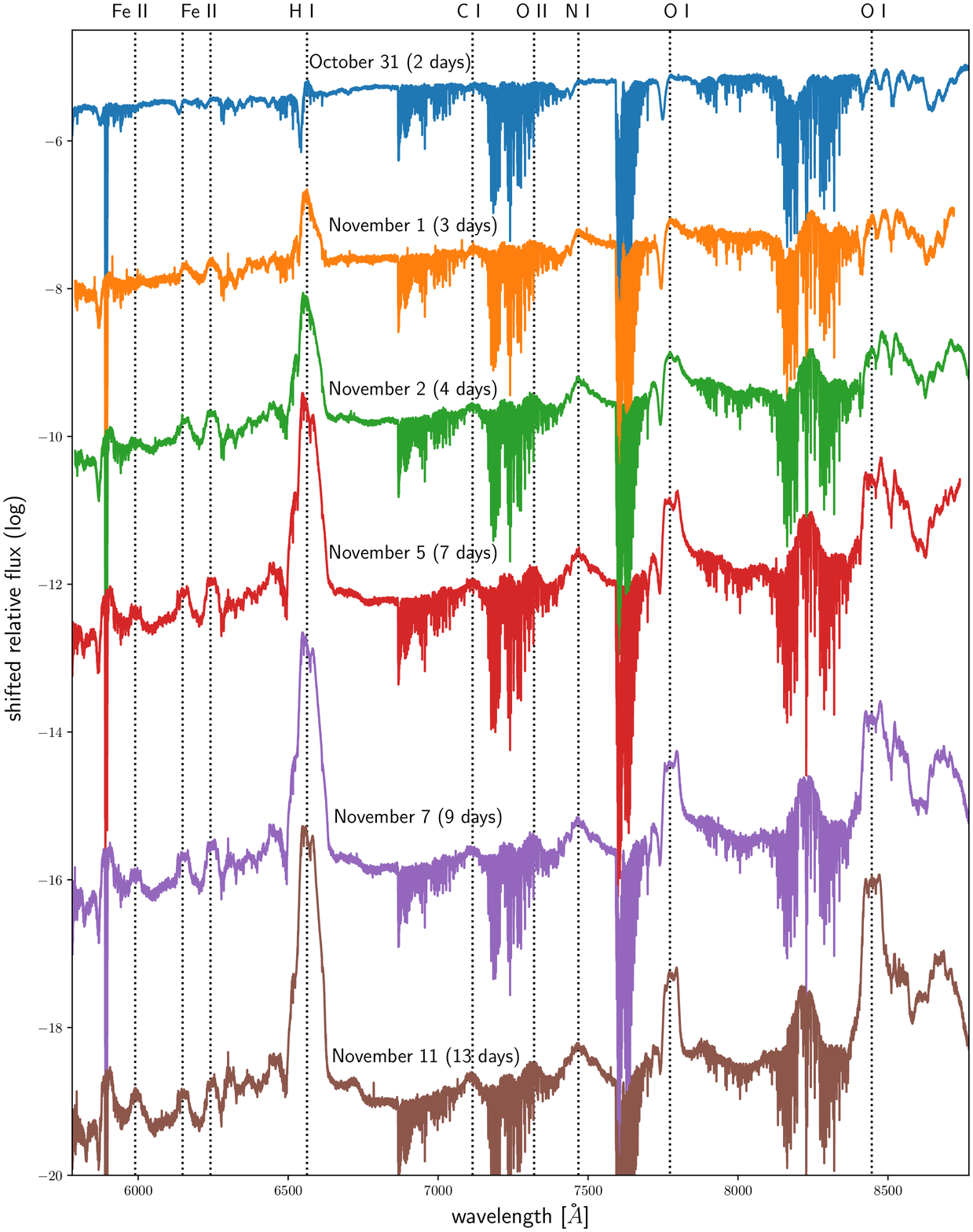}
\caption{Six spectra of Nova V659 Sct observed in the red channel
of the HEROS spectrograph. Dates are of 2019 and the corresponding days
after discovery are given in brackets. Broad bands of telluric lines are present in
this wavelength range.}
\label{fig:all_specs_R}
\end{center}
\end{figure*}

A set of six of the observed spectra of Nova V659 Sct is shown for the blue channel in Figure~\ref{fig:all_specs_B}.
The spectrum from November 8 shows very small differences to the one from November 7, and 
the spectrum from November 17 had a very low signal-to-noise ratio and no features are visible in the blue channel.
For those reasons we here present only six spectra.
The presented spectra highlight the different phases of the nova outburst. 
The first spectrum is still in the absorption phase
and was taken during the maximum of the $V$~band light curve. The next spectrum, taken one night later, shows
P-Cygni profile lines. The later spectra show the evolution towards the emission line phase.
We marked the most important features in the Nova V659 Sct spectra with vertical lines at their rest wavelengths.
One can clearly see the blueshifted absorption features of each line, while the emission features
are centered around the rest wavelengths.
In the blue channel, we clearly identify features of Ca~II, Fe~II and H~I. 
Since the Nova was close to the Sun on the sky, we were not
able to make long exposures right after sunset also with the nova being hardly above the horizon.
As a result, several wavelength regions of the blue part of some of the spectra are affected by noise and the lines are not always visible.

Six observed TIGRE spectra in the red channel are presented in Figure~\ref{fig:all_specs_R}. 
The spectra have a higher signal-to-noise ratio (S/N) than in the blue channel. However, in this wavelength range
some regions of the spectra are contaminated by the broad telluric absorption bands. 
The red spectra contain the common features of Fe~II and H~I as well as some lines
of O~I, O~II and N~I.
Here, the first spectrum shows already P-Cygni profiles. This may have to do with the spectral energy distribution
since bluewards of the maximum flux the features are often in absorption, while redwards it is P-Cygni profiles or emission.
The strongest features are the H$\alpha$ line and the two lines of O~I at 7773.0~\AA\ and 8446.3~\AA, 
which show asymmetric emission during the later phase of the spectral evolution.
The narrow and deep absorption feature around 5900~\AA\ is interstellar absorption of the Na~I doublet lines.

%
% Spectral analysis
%
\section{Spectral analysis}

Having eight TIGRE spectra of Nova V659 Sct with an intermediate resolution allows us to analyse them in detail and
identify and study the features present in the Nova. In addition, we are able to study the absorption features
of the interstellar medium in form of atomic lines and DIBs. We present the results in this following section.

\subsection{Line identification}

We thoroughly inspected all the above presented six optical spectra taken of Nova V659 Sct and 
identified all features in these spectra using the database of lines found in Nova V339 Del and published in \citet{novadel}.

% Table of features
\begin{table}[t]%
\centering
\caption{List of lines in the blue channel that have been identified in the six observed spectra of Nova V659~Sct.}\label{tab:linesblue}
\tabcolsep=0pt%
\begin{tabular*}{250pt}{@{\extracolsep\fill}cccccccc@{\extracolsep\fill}}
      \toprule
\textbf{Line [\AA]} & \textbf{Ion} & \textbf{Oct 31} & \textbf{Nov 1} & \textbf{Nov 2} & \textbf{Nov 5} & \textbf{Nov 7} & \textbf{Nov 11}\\
\midrule
3835.4 & H I & \checkmark  &  &  & &  & \\
3889.1 & H I &  \checkmark &   &  &  & & \\
3933.7 & Ca II  &  \checkmark &  \checkmark  &  \checkmark &  & \checkmark & \\
3968.5 & Ca II  &  \checkmark &  \checkmark  &  \checkmark &  & \checkmark & \\ 
3970.1 & H I  & \checkmark &  \checkmark  &  \checkmark & & \checkmark & \\ 
4101.7 & H I &  \checkmark &  &  &  & \checkmark  & \checkmark \\
4130.9 & Si II  &  \checkmark &  &  &  &  & \\
4173.5 & Fe II  &  \checkmark &  & \checkmark &  & \checkmark  & \checkmark  \\ 
4178.9 & Fe II  &  \checkmark &  & \checkmark &  & \checkmark  & \checkmark  \\
4233.2 & Fe II  &  \checkmark &  &  &  & \checkmark   &\checkmark  \\
4296.5 & Fe II & \checkmark &  &  &  & \checkmark  &\checkmark  \\
4340.5 & H I  & \checkmark & \checkmark  & \checkmark & \checkmark & \checkmark   &\checkmark  \\
4385.4 & Fe II  & \checkmark  & \checkmark  &  &  &  & \\
4416.3 & Fe II  & \checkmark  & \checkmark  & \checkmark &  &  & \\
4416.8 & [Fe II]  & \checkmark  & \checkmark  &  \checkmark &  &  & \\
4443.8 & Ti II  & \checkmark  &  &  &  &  & \\
4481.2 & Mg II  & \checkmark &  &  &  &  &  \\
4522.6 & Fe II  & \checkmark &  &  &  &  &\\
4555.8 & Fe II  & \checkmark &  &  &  &  & \\
4583.8 & Fe II  & \checkmark & \checkmark  & \checkmark & \checkmark  & \checkmark   & \checkmark \\
4629.3 & Fe II  & \checkmark &  \checkmark & \checkmark & \checkmark  & \checkmark  & \checkmark  \\
4670.4 & Sc II & \checkmark &  &  &  &  & \\ 
4772.1 & [Fe II]  & \checkmark &  &  &  &  & \\
4805.1 & Ti II  & \checkmark &  &  &  &  & \\
4861.3 & H I  & \checkmark & \checkmark & \checkmark & \checkmark  & \checkmark & \checkmark  \\
4923.9 & Fe II  & \checkmark & \checkmark & \checkmark & \checkmark  &  \checkmark & \checkmark \\
5018.4 & Fe II  & \checkmark  &  \checkmark  & \checkmark & \checkmark  & \checkmark  & \checkmark  \\
5129.2 & Ti II  & \checkmark  &  & & & &\\
5169.0 & Fe II  & \checkmark  & \checkmark  & \checkmark & \checkmark  & &\\
5197.6 & Fe II  & \checkmark  & \checkmark  & & & &\\
5234.6 & Fe II  & \checkmark  &  \checkmark & \checkmark & & & \\
5276.0 & Fe II  & \checkmark  & \checkmark  & \checkmark & \checkmark  &  \checkmark & \checkmark \\
5316.6 & Fe II  & \checkmark  & \checkmark  & \checkmark & \checkmark  &  \checkmark & \checkmark \\
5362.8 & Fe II  & \checkmark  &  & & & & \\
5425.0 & Fe II  & & & \checkmark & \checkmark  & & \\
5532.1 & Fe II  & \checkmark  & \checkmark  &  \checkmark  & \checkmark & \checkmark & \checkmark  \\
\bottomrule
\end{tabular*}
\end{table}

% second table
% Table of features
\begin{table}[t]%
\centering
\caption{List of lines in the red channel that have been identified in the six observed spectra of Nova V659~Sct.}\label{tab:linesred}
\tabcolsep=0pt%
\begin{tabular*}{250pt}{@{\extracolsep\fill}cccccccc@{\extracolsep\fill}}
      \toprule
\textbf{Line [\AA]} & \textbf{Ion} & \textbf{Oct 31} & \textbf{Nov 1} & \textbf{Nov 2} & \textbf{Nov 5} & \textbf{Nov 7} & \textbf{Nov 11}\\
\midrule
5875.6 & He I  &   &   & &  \checkmark &  \checkmark  \\
5889.9 & Na I  &   \checkmark & \checkmark  & \checkmark  & \checkmark & \checkmark  & \checkmark  \\
5895.9 & Na I  &   \checkmark & \checkmark  & \checkmark  & \checkmark & \checkmark  & \checkmark \\
5991.4 & Fe II  &   &   & &  \checkmark &  \checkmark &  \checkmark  \\ 
6148.0 & Fe II &    \checkmark & \checkmark   &  \checkmark  & \checkmark & \checkmark & \checkmark \\
6240.6 & Fe II  &   \checkmark & \checkmark   &  \checkmark   & \checkmark  &  \checkmark & \checkmark \\ 
6247.6 & Fe II  &   \checkmark &  \checkmark  &   \checkmark  & \checkmark  & \checkmark& \checkmark \\
6300.3 & [O I]  &  \checkmark & \checkmark & \checkmark  & \checkmark & \checkmark  &  \checkmark \\ 
6347.1 & Si II  &  \checkmark &  &  & & & \\ 
6371.4 & Si II  & \checkmark  & \checkmark &  \checkmark   &  \checkmark  &  \checkmark & \checkmark \\
6456.4 & Fe II  & \checkmark  & \checkmark &   &  &   & \\
6483.8 & Ni II  & \checkmark  & \checkmark &   &  &   & \\
6562.7 & H I   & \checkmark  & \checkmark &  \checkmark & \checkmark &  \checkmark & \checkmark \\
6678.2 & He I  &   & \checkmark  & \checkmark & \checkmark & \checkmark &\checkmark \\
6722.6 & N I   & \checkmark  &  \checkmark &   &  &  \checkmark & \checkmark \\
7115.0 & C I  & \checkmark & \checkmark & \checkmark  & \checkmark  &  \checkmark  & \checkmark \\
7320.0 & [O II] &  &  & \checkmark  & \checkmark  & \checkmark & \checkmark \\  
7442.3 & N I &  \checkmark  & \checkmark   &   & &   & \\
7468.2 & N I &  \checkmark  & \checkmark   & \checkmark  &  \checkmark  &  \checkmark & \checkmark\\
7773.0 & O I & \checkmark  & \checkmark & \checkmark & \checkmark & \checkmark & \checkmark\\
8220.0 & Fe II  & \checkmark  & \checkmark & \checkmark & \checkmark & \checkmark & \checkmark\\
8446.3 & O I & \checkmark  &  \checkmark & \checkmark &  \checkmark & \checkmark  & \checkmark \\
8498.0 & Ca II & \checkmark  &  \checkmark  &  \checkmark &  &  & \\
8542.1 & Ca II & \checkmark  &  \checkmark & \checkmark &  &  & \\
8598.4 & H I & \checkmark  &  &  &  &  & \\
8662.1 & Ca II & \checkmark  &  \checkmark &  \checkmark  &  &  & \\
8665.0 & H I & \checkmark  & \checkmark &  \checkmark  & &  & \\
8750.5 & H I & \checkmark  &  &  &  &  & \\
\bottomrule
\end{tabular*}
\end{table}

We present the list of lines that we identified in the blue channel of the HEROS spectrograph 
of Nova V659 Sct in Table~\ref{tab:linesblue}.
Since the spectrum from November 8 is almost the same as the one from November 7 the identified
lines in the spectrum from November 7 are also visible in the spectrum from November 8.
The spectrum from November 17 is only noise in the blue channel, and no lines are visible.
Below $4400$~\AA\ some of the spectra have a very low signal-to-noise ratio, because of the low flux and the relatively short
exposure times due to the short observable time of the Nova on the sky right after sunset.
Therefore, some features probably do not disappear
but are just not clearly visible above the strong noise. Especially, the second spectrum from November 1, had a very low
S/N due to bad weather conditions.
The spectra of Nova V659 Sct contain many Fe~II lines. The common hydrogen lines like H$\beta$ are clearly visible. 
In addition, we could identify lines of Ca~II, Si~II, Ti~II and Mg~II. We also identified two forbidden lines. 
In summary, some lines disappear quickly during the spectral evolution, while
the strong lines are visible in every spectrum up to the emission line phase.

The features we could identify in the red channel of the observed spectra of Nova V659 Sct are presented in Table~\ref{tab:linesred}.
Apart from the prominent H$\alpha$ line and the strong O~I feature at 7773.0~\AA, we found in the red channel spectra 
many lines of Fe~II. In addition, we could identify lines of O~II, Na~I, Si~II, Ni~II, Ca~II, C~I and N~I.
We also identified two forbidden lines.
In the later spectra the features of the Ca~II triplet and the H~I line around 8500~\AA\ are strongly blended and one cannot
distinguish and clearly identify the contribution from each one of these lines to the broad emission feature observed
at that wavelength.
There could also be some features hidden in the strong absorption bands of telluric lines present in the red channel
of the spectra of Nova V659 Sct.
In the low S/N spectrum from November 17 only three features of the H$\alpha$ line and two O~I lines
at 7773.0 and 8446.3~\AA\ are visible.

\subsection{Spectral evolution of prominent features}

\begin{figure}
\begin{center}
 \resizebox{\hsize}{!}{\includegraphics{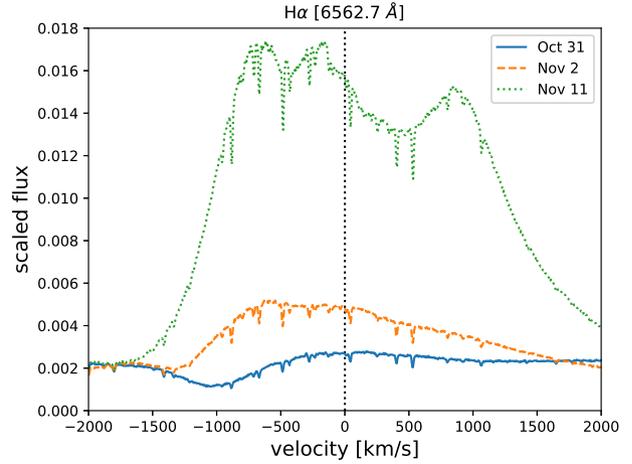}}
\caption{Spectral evolution of the H$\alpha$ feature. The narrow lines come from telluric absorption.}
\label{fig:halpha}
\end{center}
\end{figure}

\begin{figure}
\begin{center}
 \resizebox{\hsize}{!}{\includegraphics{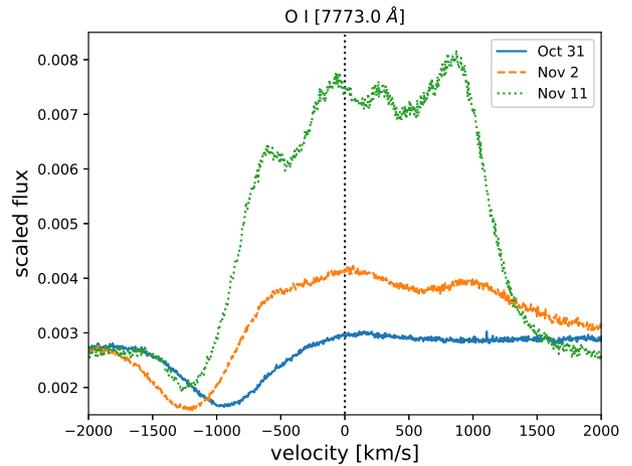}}
\caption{Spectral evolution of O~I feature at 7773.0~\AA. }
\label{fig:oi}
\end{center}
\end{figure}

To present some examples for the spectral evolution of the features in the spectra of Nova V659 Sct, in Figure~\ref{fig:halpha}, we show three spectra of the prominent H$\alpha$ line.
The first spectrum from October 31 shows a clear absorption feature around $-1000$~km~s$^{-1}$. Two days later, the absorption feature has moved bluewards to higher 
expansion velocities. The emission feature starts to show up around the rest wavelength. The last spectrum from November 11 shows the H$\alpha$ line completely in emission. 
The emission feature is asymmetric and has two peaks on the blue side and one peak at the red part of the feature. The rightmost peak is a bit lower than
the two peaks on the left.
 
The spectral evolution of the prominent O~I line at 7773.0~\AA\ is shown in Figure~\ref{fig:oi}. The spectral evolution is similar to the H$\alpha$ line as the first spectrum
shows a strong absorption feature around $-1000$~km~s$^{-1}$.
This absorption feature moves towards higher expansion velocities and an emission feature starts to appear. The emission feature of the last spectrum from November 11 is also asymmetric. 
However, the peak on the red side is the highest, in contrast to the H$\alpha$ line. Another difference is that the absorption feature does not disappear.

\begin{table*}[t]%
\centering
\caption{Expansion velocities of the absorption features of H$\alpha$, H$\beta$, the O~I line at 7773.0~\AA\ line and two Fe~II lines at 5018.4 and 5196.0~\AA.}
\tabcolsep=0pt%
\begin{tabular*}{450pt}{@{\extracolsep\fill}lccccc@{\extracolsep\fill}}
      \toprule
      \textbf{Date} & \textbf{$v_\mathrm{H\alpha}$ [km~s$^{-1}$]} & \textbf{$v_\mathrm{H\beta}$ [km~s$^{-1}$]}  & \textbf{$v_\mathrm{O\;I,\;7773}$ [km~s$^{-1}$]} & \textbf{$v_\mathrm{Fe\;II,\;5018}$ [km~s$^{-1}$]}  & \textbf{$v_\mathrm{Fe\;II,\;5169}$ [km~s$^{-1}$]}\\
      \midrule
      \textbf{Oct 31}  &  $-1052 \pm 23$ &  $-962\pm37$  &  $ -931\pm 10$ &  $-986\pm23$  &  $-980\pm23$ \\ 
      \textbf{Nov 01}  &  $-1255 \pm 45$ & $-1258\pm62$  &  $-1174\pm 19$ & $-1230\pm48$  & $-1189\pm59$ \\ 
      \textbf{Nov 02}  &  $-1371 \pm 17$ & $-1326\pm24$  &  $-1280\pm 11$ & $-1308\pm48$  & $-1270\pm35$ \\
      \textbf{Nov 05}  &  $-1522 \pm 78$ & $-1597\pm105$ &  $-1390\pm 10$ & $-1320\pm83$  & $-1334\pm41$ \\
      \textbf{Nov 07}  &                 & $-1597\pm124$ &  $-1317\pm 10$ & $-1404\pm120$ &              \\  
      \textbf{Nov 08}  &                 & $-1418\pm124$ &  $-1273\pm 10$ & $-1320\pm120$ &              \\  
      \textbf{Nov 11}  &                 &               &  $-1252\pm 10$ &               &              \\ 
      \textbf{Nov 17}  &                 &               &  $-1099\pm 10$ &               &              \\ 
 \bottomrule
\end{tabular*}
\label{vel}
\end{table*}
\begin{figure}
\begin{center}
 \resizebox{\hsize}{!}{\includegraphics{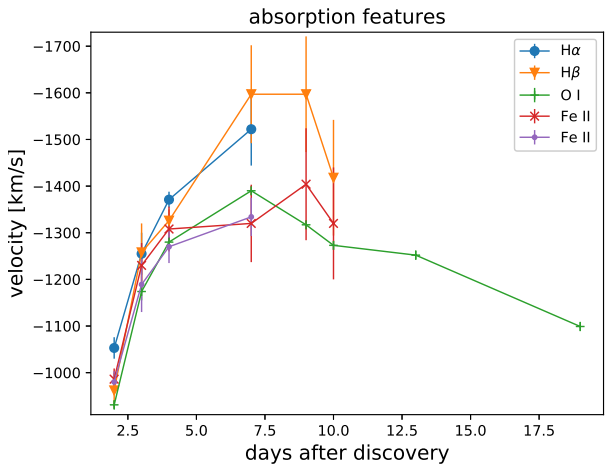}}
\caption{Evolution of the expansion velocities of the absorption features of five spectral lines that correspond
to three different species.}
\label{fig:vels}
\end{center}
\end{figure}

To quantify the spectral evolution, we measured the velocities of the absorption feature of H$\alpha$, H$\beta$ and the O~I 7773.0~\AA\ line as well
as two Fe~II lines at 5018.4 and 5169.0~\AA\ and present the results in Table~\ref{vel} and Figure~\ref{fig:vels}.
The absorption features of all lines move towards higher expansion velocities.
For the H$\alpha$ line the last three spectra do not show a minimum in the spectrum but
rather a flat part. Thus, it was not possible to determine an expansion velocity.
The absorption feature of H$\beta$ has similar expansion velocities as the H$\alpha$ line,
except for the first observation on October 31, where the H$\alpha$ absorption feature has a higher
expansion velocity than that of H$\beta$.

The velocities of the O~I line absorption feature has slightly slower expansion velocities than the H$\alpha$ absorption feature.
The absorption feature of the O~I line moves after November 5 back to smaller expansion velocities. 
However, we do not have a daily coverage around that time so the "maximum" could have occurred on a different day.
The expansion velocities of the two Fe~II lines are very similar. Nevertheless, they do not rise as high as the ones
of the hydrogen lines. On November 5, the expansion velocities of the hydrogen lines are over 1500~km~s$^{-1}$, while
the ones of the lines of Fe~II and O~I are lower than 1400~km~s$^{-1}$.

In addition, we measured the positions of the emission peaks of the H$\alpha$ line in the spectrum obtained
on November 11, about 13 days after the discovery to determine their velocities.
We found that the bluemost emission peak is located at a velocity of $-620\pm10$~km~s$^{-1}$. The middle peak, which is still
a bit shifted to the blue, has a velocity of $-177\pm14$~km~s$^{-1}$. The peak at the red part of the H$\alpha$ emission feature
is located at a velocity of $845\pm18$~km~s$^{-1}$.

In comparison, the O~I emission feature has its highest peak when the H$\alpha$ feature has its lowest. 
In total, the O~I 7773.0~\AA\ line has four peaks. The bluemost peak, which is also the lowest,
has a velocity of $-576\pm12$~km~s$^{-1}$. The velocity is similar to that of the blue peak of the H$\alpha$ emission feature.
For the second peak we determined a velocity of $-73\pm20$~km~s$^{-1}$. The third peak is already redshifted and
presents a velocity of $279\pm50$~km~s$^{-1}$.
The redmost peak has a velocity of about $868\pm 30$~km~s$^{-1}$.
The velocity of the red peak is practically the same as that of the respective peak of the H$\alpha$ line. 
However, as already mentioned, the form of the emission feature is the opposite.

It is important to mention that the shape of the H$\beta$ emission feature is similar to the one of the H$\alpha$ feature. 
The highest peak is also located on the blue side of the feature. It has also two peaks on the blue side and one on the red side of the feature. 
We found that the Fe~II emission features are mainly flat without any significant asymmetries or peaks.

\begin{table*}[t]%
\centering
\caption{Expansion velocities of the rightmost emission features of H$\alpha$, H$\beta$ and the two O~I lines at 7773.0 and 8446.3~\AA.}
\tabcolsep=0pt%
\begin{tabular*}{350pt}{@{\extracolsep\fill}lcccc@{\extracolsep\fill}}
      \toprule
      \textbf{Date} & \textbf{$v_\mathrm{H\alpha}$ [km~s$^{-1}$]} & \textbf{$v_\mathrm{H\beta}$ [km~s$^{-1}$]}  & \textbf{$v_\mathrm{O\;I,\;7773}$ [km~s$^{-1}$]} & \textbf{$v_\mathrm{O\;I,\;8446}$ [km~s$^{-1}$]} \\
      \midrule
      \textbf{Nov 05}  &  $927 \pm 30$ & $956 \pm 21$ &  $964\pm 23$ & $1086 \pm 53$ \\
      \textbf{Nov 07}  &  $854 \pm 23$ & $882 \pm 30$ &  $887\pm 19$ & $1008 \pm 18$ \\  
      \textbf{Nov 08}  &  $854 \pm 23$ & $857 \pm 19$ &  $890\pm 19$ & $ 923 \pm 18$ \\  
      \textbf{Nov 11}  &  $845 \pm 18$ & $826 \pm 24$ &  $868\pm 30$ & $ 870 \pm 30$ \\ 
      \textbf{Nov 17}  &  $831 \pm 23$ &              &  $825\pm 70$ & $ 841 \pm 70$ \\ 
 \bottomrule
\end{tabular*}
\label{vel_em}
\end{table*}
\begin{figure}
\begin{center}
 \resizebox{\hsize}{!}{\includegraphics{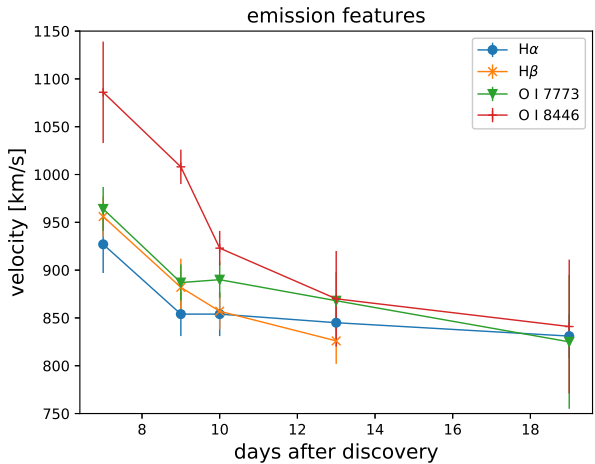}}
\caption{Evolution of the expansion velocities of the rightmost emission features of four spectral lines that correspond
to two different species.}
\label{fig:vels_em}
\end{center}
\end{figure}

The rightmost feature of the emission lines is the clearest and also clearly distinguishable in most of the lines.
Thus, we measured the expansion velocity of the peak of these features for the lines of H$\alpha$, H$\beta$ and two O~I lines at
7773.0~\AA\ and 8446.3~\AA\ and present the results in Table~\ref{vel_em}. 
As Figure~\ref{fig:vels_em} illustrates, the expansion velocity indicated by this feature decreases with time in all lines.
The expansion velocity of the rightmost emission feature of the O~I line at 8664.3~\AA\ is the highest with values larger than 1000~km~s$^{-1}$,
while the other features show very similar expansion velocities.

%
%  ISM
%
\subsection{Interstellar absorption features}

\subsubsection{Atomic absorption}

The intermediate resolution spectra of Nova V659 Sct show several absorption features
of the interstellar medium (ISM). We added up the flux of all our eight TIGRE spectra in order
to obtain one spectrum with a higher S/N to be able to identify and measure the ISM absorption features.
We found clear absorption features for the common atomic IS lines of Ca~II and Na~I.
There is indication of both common K~I lines at 7664.91 and 7698.97~\AA\ in the spectra of Nova V659 Sct,
but we cannot analyse these features since the lines are strongly blended with telluric absorption lines.

\begin{figure}
\begin{center}
 \resizebox{\hsize}{!}{\includegraphics{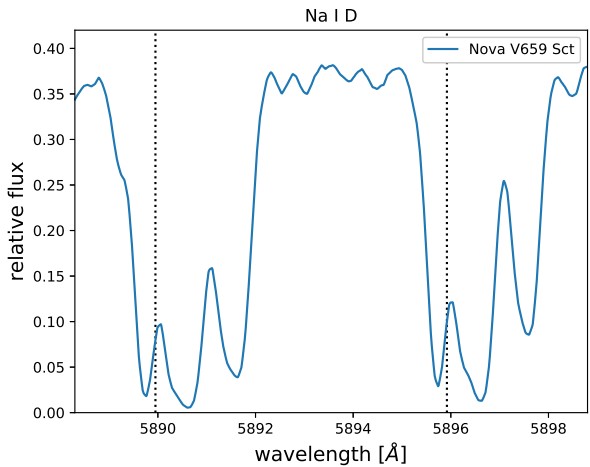}}
\caption{Interstellar absorption features of Na~I in Nova V659~Sct. Three clear subfeatures are visible.}
\label{fig:nai}
\end{center}
\end{figure}

The IS absorption of Na~I presented in Figure~\ref{fig:nai} shows three strong and clear subfeatures.
The strongest feature in the middle seems to have another subfeature at its left side, but it is not clearly distinguishable. 
The vertical lines represent the rest wavelengths of the two Na~I D doublet lines. One absorption feature is
slightly blueshifted, while the other two features are shifted to the red.

\begin{table}[t]%
\centering
\caption{Velocities and equivalent widths (EW) of the atomic interstellar absorption features of Ca~II and Na~I found in Nova V659 Sct.}
\tabcolsep=0pt%
\begin{tabular*}{250pt}{@{\extracolsep\fill}lccc@{\extracolsep\fill}}
      \toprule
      \textbf{Ion} & \textbf{wavelength [\AA]} & \textbf{velocity [km~s$^{-1}$]} & \textbf{EW [m\AA]}\\
      \midrule
Ca~II & 3933.66  &  $21.4\pm2.3$  & $1649  \pm 100$\\
Ca~II & 3968.47  &  $22.1\pm2.5$  & $1014  \pm 55$ \\
Na~I  & 5889.95  &  $-9.5\pm0.6$  & $641  \pm 40$\\
      &          &  $30.3\pm0.4$  & $1315 \pm 100$\\
      &          &  $84.1\pm0.5$  & $599  \pm 60$\\
Na~I  & 5895.92 &  $-10.1\pm0.6$  & $493  \pm 20$\\
      &          &  $30.0 \pm0.6$  & $1098 \pm 100$\\
      &          &  $86.8\pm0.6$  & $440 \pm 30$\\
 \bottomrule
\end{tabular*}
\label{atomic}
\end{table}

We determined the minimum of the absorption features and measured their equivalent widths (EW).
We present the results of our measurements in Table~\ref{atomic}. The velocities of the respective features
of the two Na~I D lines are similar.
The features of the Ca~II H \& K lines suffer from a very low S/N. There seems to be a substructure at the right part of the feature.
However, we cannot distinguish this substructure and, therefore, determined the velocity and the equivalent width assuming just one feature.
The velocity of the Ca~II absorption features is comparable to the one of the middle feature of the Na~I lines.
There exists also interstellar absorption of K~I. However, these features are strongly blended with telluric lines, and it is
impossible to perform any measurements. 

\subsubsection{Diffuse interstellar bands}

\begin{table}[t]%
\centering
\caption{Identified DIBs in the spectra of Nova V659 Sct and their respective velocities and equivalent widths (EW).}
\tabcolsep=0pt%
\begin{tabular*}{250pt}{@{\extracolsep\fill}lcc@{\extracolsep\fill}}
      \toprule
      \textbf{Feature}& \textbf{velocity [km~s$^{-1}$]} & \textbf{EW [m\AA]}\\
      \midrule
      DIB 5780  &  $19.2 \pm 7.8$ &  $602\pm23$\\ 
      DIB 5797  &  $-6.7 \pm 5.2$ &  $238\pm7$\\ 
      DIB 5850  &  $-1.0 \pm 4.6$ &  $69 \pm2$\\
      DIB 6196  &  $2.1  \pm 4.8$ &  $52 \pm2$\\
      DIB 6203  &  $31.6 \pm 7.3$ &  $276 \pm 4$\\  
      DIB 6379  &  $-14.7\pm 1.5$ &  $69 \pm 1$\\ 
      DIB 6614  &  $8.6  \pm 4.1$ &  $238 \pm 8$\\
      DIB 6660  &  $-16.3\pm 2.9$ &  $22 \pm 1$\\
      DIB 7562  &  $-5.1 \pm 6.4$ &  $171 \pm 3$\\
      DIB 7581  &  $-11.1\pm 7.1$ &  $36 \pm 2$\\
 \bottomrule
\end{tabular*}
\label{dibs}
\end{table}

The intermediate resolution spectra of Nova V659 Sct contain several features of diffuse interstellar bands (DIB). 
We measured the equivalent widths of the features by integrating over the features. In addition, we determined
the position of the minima in order to estimate the velocity of the respective DIB features.
We used for that the rest wavelengths for the DIBs given in \citet{hobbs09}.
Inspecting the spectra of Nova V659 Sct, we could identify several DIBs. 
The measured velocities and equivalent widths (EW) of the features are given in Table~\ref{dibs}.
DIB 5780 is the strongest DIB with the highest EW that we found. 
All of the DIBs found in the spectra of Nova V659 Sct were also present in the spectra the other
two novae that we observed with the TIGRE telescope \citep{jack19}.
There are two DIBs (5850 and 7581) that are "new" in Nova V659 Sct. 
In general, the EWs of the DIBs are significantly higher in the spectra of Nova V659 Sct, which agrees with the
high galactic reddening of $E(B-V)=0.9$ found by \citet{swift}.
The values for the velocities are somewhat distributed over a larger range from $-14.7$ to 31.6~km~s$^{-1}$. 
However, the S/N of the spectra was not very high, and it was difficult to determine the position of the minimum
and, thus, the velocity of the DIB features.

%
% Summary
%
\section{Summary and Conclusions}

We obtained a series of eight optical spectra (3,800 to 8,800~\AA) of the Nova V659 Sct during
the different phases of its outburst with
our robotic 1.2~m TIGRE telescope and its intermediate resolution ($R\approx 20,000$) HEROS spectrograph.
Using the MMRD relation we could place the Nova at a distance of about 7.5~kpc.
A thorough line identification was performed finding lines of H~I, Fe~II, O~I, Na~I and Ca~II, among others.
The absorption features of the H$\alpha$ and the O~I 7773.0~\AA\ lines 
move during the first days of the spectral evolution to higher expansion velocities.
The emission features are asymmetric with three peaks for the H$\alpha$ line or four peaks for the O~I line. 
While in the H$\alpha$ line the rightmost peak was the lowest, it was the strongest in the O~I line.
The position of the outermost peaks are about the same.
The expansion velocity of the rightmost emission feature is decreasing in time.

This Nova went from optically thick absorption line spectrum into optically thin nebular phase very
quickly, indicating a small envelope mass. This is supposed to be typical for a more evolved system
with a relatively large WD mass, as such erupting more frequently with only small amounts of accreted
hydrogen igniting easily.
In comparison with the Novae V339 Del and V5668 Sgr, the absorption features of the spectral lines in V659 Sct show
higher expansion velocities. 

Thanks to the resolution of the spectra of Nova V659 Sct, we were able to study the absorption features of the interstellar medium.
Concerning the atomic ISM absorption features, we found that both sodium D lines show a substructure with three main components. 
Interstellar absorption of Ca~II is present, but it is not possible
to distinguish any substructures due to the low S/N in that part of the spectra. 
We also identified the features of both K~I lines, but they are strongly blended with telluric absorption lines.
Several DIBs were identified in the spectra of Nova V659 Sct, even two more than in Novae V339 Del and V5668 Sgr.
We determined the velocities and the equivalent widths of all DIB features. The DIBs are relatively strong due
to high galactic reddening (e.g. absorption), and this is confirmed by the high value of $E(B-V)=0.9$
mentioned earlier.

\section*{Acknowledgments}

The authors are grateful for financial support by the joint bilateral project CONACyT-DFG No. 278156.

We acknowledge with thanks the variable star observations from the AAVSO International Database contributed by
observers worldwide and used in this research.

\bibliography{all}

\appendix

\section{Observation details}
\begin{table*}[t]%
\centering
\caption{Details of the TIGRE observations.}
\tabcolsep=0pt%
\begin{tabular*}{380pt}{@{\extracolsep\fill}lcccccc@{\extracolsep\fill}}
      \toprule
      \textbf{Date} & \textbf{Time} & \textbf{Julian Date}  & \textbf{Airmass} & \textbf{Avg. Seeing} & \textbf{Exposure} & \textbf{Mean S/N} \\
      \textbf{(UT)} & \textbf{(UT)} & \textbf{}  & \textbf{} & \textbf{[arcsec]} & \textbf{[sec]} & \textbf{} \\
       \midrule
      \textbf{Oct 31}  &  $00:52:24$ & $2458787.536389$ & $1.375994$ & $1.440504$ & $1800$ & $90.2$ \\ 
      \textbf{Nov 01}  &  $00:51:44$ & $2458788.535929$ & $1.389399$ & $3.494088$ & $3600$ & $42.7$ \\ 
      \textbf{Nov 02}  &  $00:50:56$ & $2458789.535371$ & $1.402655$ & $1.983384$ & $1800$ & $63.0$ \\
      \textbf{Nov 05}  &  $01:21:39$ & $2458792.556700$ & $1.63851$  & $1.524744$ & $3600$ & $55.3$ \\
      \textbf{Nov 07}  &  $00:48:45$ & $2458794.533852$ & $1.48339$  & $2.543112$ & $3600$ & $47.0$ \\  
      \textbf{Nov 08}  &  $00:48:48$ & $2458795.533888$ & $1.505066$ & $1.689012$ & $3600$ & $60.1$ \\  
      \textbf{Nov 11}  &  $00:47:06$ & $2458798.532709$ & $1.564179$ & $1.302912$ & $3600$ & $47.3$ \\ 
      \textbf{Nov 17}  &  $00:59:45$ & $2458804.541489$ & $1.858462$ & $1.798992$ & $1800$ & $9.6$  \\ 
 \bottomrule
\end{tabular*}
\label{obs_log}
\end{table*}

We present the details of our TIGRE observations in Table~\ref{obs_log}. The start of the observations is given in UT. 
The exposure times were either 30 minutes or one hour. The mean S/N is taken from the spectrum of the red channel.

\end{document}